\documentclass[aps,prl,twocolumn,amsmath,amsfonts,amssymb,nofootinbib,
  secnumarabic,notitlepage,superscriptaddress]{revtex4-2}
\usepackage[utf8]{inputenc}
\usepackage{amsmath,amsfonts,amssymb,graphicx,subfigure}
\usepackage{slashed}
\usepackage{multirow}
\usepackage{lineno}
\usepackage{array}
\usepackage[dvipsnames]{xcolor}
\usepackage[colorlinks=true, linkcolor = blue]{hyperref} 
\hypersetup{
pdftitle={Probing the Weinberg Operator at Colliders -- [arXiv:2012.09882] --     Phys. Rev. D103 (2021) 115014}, pdfauthor={Fuks, et al}, 
colorlinks=true, linkcolor=ForestGreen, citecolor=ForestGreen,
filecolor=ForestGreen, urlcolor=ForestGreen}
\graphicspath{{Plots/}}
\def\ie{{\it i.e.}}

\newcommand{\be}{\begin{equation}}
\newcommand{\ee}{\end{equation}}

\def\lag{{\cal L}}

\newcommand{\nuBB}{{0\nu\beta\beta}}

\newcommand{\MeV}{{\rm ~MeV}}
\newcommand{\GeV}{{\rm ~GeV}}
\newcommand{\TeV}{{\rm ~TeV}}

\newcommand{\fb}{{\rm ~fb}}
\newcommand{\ab}{{\rm ~ab}}
\newcommand{\invfb}{{\rm ~fb^{-1}}}
\newcommand{\invab}{{\rm ~ab^{-1}}}
\newcommand{\etmiss}{{E_{\rm T}^{\rm miss}}}
\newcommand{\libName}{{\sc\small SMWeinberg}}

\newcommand{\fr}{{\sc \small FeynRules}}

\newcommand{\confirm}[1]{{\color{black}#1}}

\begin{document}
\leftline{}
\rightline{CP3-20-63, DESY 20-230, IFJPAN-IV-2021-3, MCNet-20-23}
\title{Probing the Weinberg Operator at Colliders}

\author{Benjamin~Fuks}
\email{fuks@lpthe.jussieu.fr}
\affiliation{Sorbonne Universit\'e, CNRS, Laboratoire de Physique Th\'eorique et
  Hautes \'Energies, LPTHE, F-75005 Paris, France}
\affiliation{Institut Universitaire de France, 103 boulevard Saint-Michel,
  75005 Paris, France}

\author{Jonas Neundorf}
\email{jonas.neundorf@desy.de}
\affiliation{Deutsches Elektronen-Synchrotron, Notkestraße 85, 22607 Hamburg, Germany}

\author{Krisztian Peters}
\email{krisztian.peters@desy.de}
\affiliation{Deutsches Elektronen-Synchrotron, Notkestraße 85, 22607 Hamburg, Germany}

\author{Richard Ruiz}
\email{rruiz@ifj.edu.pl}
\affiliation{Institute of Nuclear Physics, Polish Academy of Sciences, ul. Radzikowskiego, Cracow 31-342, Poland}
\affiliation{Centre for Cosmology, Particle Physics and Phenomenology {\rm (CP3)},\\
Universit\'e Catholique de Louvain, Chemin du Cyclotron, Louvain la Neuve, B-1348, Belgium}

\author{Matthias Saimpert}
\email{matthias.saimpert@cern.ch}
\affiliation{CERN - 1211 Geneva 23 - Switzerland}

\begin{abstract}
Motivated by searches for  $0\nu\beta\beta$ decay in nuclear experiments and collider probes of lepton number violation at dimension $d\geq7$, we investigate the sensitivity to the  $d=5$  Weinberg operator using the non-resonant signature $pp\to \ell^\pm \ell'^{\pm} j j$ at the LHC. We develop a prescription for the  operator that is applicable in collisions and decays, and focus on the $\ell\ell'=\mu\mu$ channel, which is beyond the reach of nuclear decays. For a Wilson coefficient $C^{\mu\mu}_5=1$, scales as heavy as $\Lambda\sim 8.3~(11)$~TeV can be probed with $\mathcal{L}=300~{\rm fb}^{-1}~(3~{\rm ab}^{-1})$. This translates to an effective $\mu\mu$ Majorana mass of $\vert m_{\mu\mu}\vert\sim7.3~(5.4)$~GeV, and establishes a road map for testing the Weinberg operator at accelerators.
\end{abstract}

\date{\today}

\maketitle


\textbf{\textit{Introduction}} -- Among the most pressing mysteries shared in cosmology, nuclear, and high-energy physics is whether neutrinos are their own antiparticles~\cite{Strategy:2019vxc,EuropeanStrategyGroup:2020pow}. This importance follows from Majorana neutrinos being necessary ingredients for standard leptogenesis, grand unification, as well as new gauge symmetries. Discovering that neutrinos are Majorana particles would indicate that lepton number (LN) symmetries are not conserved below the electroweak (EW) scale, and demonstrate the existence of a mass-generating mechanism beyond those responsible for chiral and EW symmetry breaking (EWSB). 

Motivated by this, broad, complementary approaches are taken to explore the nature of neutrinos~\cite{Atre:2009rg,Rodejohann:2011mu,Bilenky:2014uka,Deppisch:2015qwa,Cai:2017jrq,Cai:2017mow,Cirigliano:2018yza,Dolinski:2019nrj}. A foremost probe is the search for the neutrinoless $\beta\beta$ process $(\nuBB)$ in decays of nuclei. This is characterized by the  transition $(A,Z)\to (A,Z+2)$ and the appearance of two same-sign electrons but an absence of neutrinos in the final state. While no discovery has been confirmed, and assuming that the decay is mediated solely by the light neutrinos observed in nature, searches  place upper limits of \confirm{$79-180$ meV} at 90\% confidence level (CL)~\cite{Agostini:2020xta} on the so-called effective $\beta\beta$ Majorana mass, given by~\cite{KlapdorKleingrothaus:2000gr}
\begin{equation}
\vert m_{ee} \vert = \left\vert \sum_{k=1}^3  U_{e k} m_{\nu_k} U_{e k} \right\vert.
 \label{eq:effMajMass}
 \end{equation}
In this definition, $m_{\nu_k}$ are the mass eigenvalues of the three light neutrinos and $U_{\ell k}$ are the  Pontecorvo-Maki-Nakagawa-Sakata (PMNS) mixing matrix elements.
 
 From the perspective that the Standard Model (SM)  of particle physics is a low-energy effective field theory (EFT), Majorana neutrino masses, and hence $\vert m_{ee}\vert$, can be generated most minimally~\cite{Ma:1998dn,Bonnet:2012kz}  at dimension $d=5$ from the LN-violating Weinberg operator~\cite{Weinberg:1979sa}:
 \begin{equation}
 \lag_5 = \ \frac{C_5^{\ell\ell'}}{\Lambda} \big[\Phi\!\cdot\! \overline{L}^c_{\ell }\big]
    \big[L_{\ell'}\!\!\cdot\!\Phi\big] + \text{H.c.}
    \label{eq:dim5Lag}
 \end{equation}
Here, $\Lambda$ is the scale at which the particles responsible for LN violation become relevant degrees of freedom; $C_5^{\ell\ell'}$ is a  flavor-dependent Wilson coefficient; $L_\ell^T=(\nu_\ell,\ell)$ is the  left-handed (LH) lepton doublet; and $\Phi$ is the SM Higgs doublet, whose vacuum expectation value (vev) $v=\sqrt{2}\langle\Phi\rangle\approx246\GeV$ generates the quantity
\begin{equation}
  \confirm{m_{\ell\ell'} = C_5^{\ell\ell'} v^2/\Lambda}.
\label{eq:dim5Mass}\end{equation}

As the Weinberg operator can be realized by tree- and loop-level Seesaw models~\cite{Ma:1998dn,Bonnet:2012kz,Cai:2017jrq}, limits on $ \vert m_{ee} \vert$ translate into lower bounds on the Seesaw scale of about {$(\Lambda/C_5^{ee})\gtrsim (3.3-7.6)\cdot10^{14}\GeV$}. While stringent, a caveat of this constraint is its flavor dependence. For instance, $C_5^{ee}$ can  be  zero due to a flavor symmetry~\cite{Jenkins:2008ex}, or be immeasurably small due to accidental cancellations~\cite{Asaka:2020wfo,Asaka:2020lsx,Asaka:2021hkg}. More generally, the production of same-sign leptons involving  muons or taus in $(A,Z)\to (A,Z+2)$ decays is kinematically forbidden. Their production requires higher energies, implying a lack of sensitivity for non-electron flavors of $C_5^{\ell\ell'}$ at $\nuBB$ decay experiments.

  \begin{figure}
\includegraphics[width=\columnwidth]{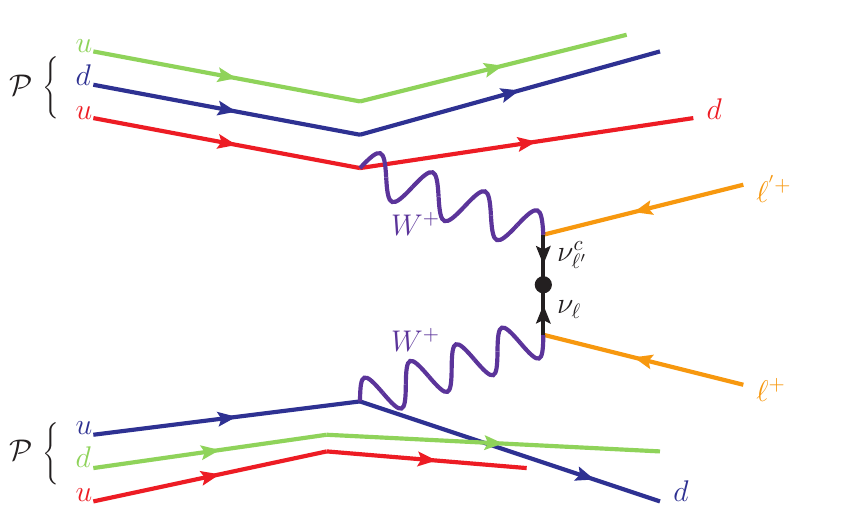}
\caption{
  Diagrammatic representation of same-sign charged lepton production through same-sign $WW$ scattering in proton collisions when mediated at dimension $d=5$.
}
\label{fig:wwScattWeinberg5_diagram}
\end{figure}

 Motivated by these limitations and by sensitivity projections for interactions at  $d\geq7$  in same-sign $W^\pm W^\pm$ scattering~\cite{Dicus:1991fk,Datta:1993nm,Aoki:2020til,Fuks:2020att},  we report an investigation into the realization of the $\nuBB$ process at \mbox{$d=5$} in high-energy proton collisions. As shown in Fig.~\ref{fig:wwScattWeinberg5_diagram}, the transition proceeds from $W^\pm W^\pm$ scattering into same-sign charged lepton pairs $\ell^\pm \ell^{'\pm}$  of arbitrary flavor and bridged via the coupling $m_{\ell\ell'}\!\propto\! C_5^{\ell\ell}/\Lambda$. While related, this work differs from studies on the ``inverse'' $\nuBB$ process~\cite{Rizzo:1982kn,Belanger:1995nh,Aoki:2020til,Fuks:2020att}, which focus on $d\geq7$ operators or their realizations. Moreover, this work relies on a new method for modeling the Weinberg operator that is applicable to meson and lepton decays, and establishes a road map to the Weinberg operator at accelerators. Finally, we release an implementation of this method in new and publicly available software\footnote{Available from \href{https://feynrules.irmp.ucl.ac.be/wiki/SMWeinberg}{feynrules.irmp.ucl.ac.be/wiki/SMWeinberg}.}  for Monte Carlo (MC) simulations.\\

\textbf{\textit{The Standard Model at Dimension Five}} -- 
To describe Majorana neutrino masses and the $\nuBB$ process from \mbox{$d=5$} interactions, we work in the SM effective field theory~\cite{Weinberg:1980wa} and  extend the SM Lagrangian $(\mathcal{L}_{\rm SM})$ by 
gauge-invariant operators of $d>4$. In the canonical representation~\cite{Grzadkowski:2010es}, the Lagrangian is given by~\cite{Weinberg:1979sa}
\begin{align}
    \mathcal{L}_{\rm SM~EFT} = \mathcal{L}_{\rm SM} + \mathcal{L}_5 + \mathcal{O}\left(\Lambda^{-2}\right),
\label{eq:lagFull}
\end{align}
where $\mathcal{L}_5$ is defined in Eq.~\eqref{eq:dim5Lag}. By the power counting of Ref.~\cite{Grzadkowski:2010es}, the Weinberg operator is the only gauge-invariant operator at $d=5$ in the SM~\cite{Weinberg:1979sa,Kobach:2016ami}.

After EWSB, the Higgs field can be expanded about its vev, which in the unitary gauge reads $\sqrt{2}\Phi\approx v+h$, where  $h$ is the  Higgs boson. The resulting Lagrangian is
\begin{align}
\mathcal{L}_5 
=
    &-\frac{C_5^{\ell\ell'}}{2\Lambda}	hh   \overline{\nu^c_\ell}\nu_{\ell'}^{}
    -\frac{C_5^{\ell\ell'}v}{\Lambda}	h\overline{\nu^c_\ell}\nu_{\ell'}^{}
   \nonumber\\
    &-
     \frac{C_5^{\ell\ell'}v^2}{2\Lambda}  \overline{\nu^c_\ell}\nu_{\ell'}^{}  + \text{H.c.}
     \label{eq:lag5_ewsb}
\end{align}
Here, $C^{\ell\ell'}_5$ is defined in the flavor basis. The minus signs above originate from the SU$(2)_L$-invariant product $\Phi\cdot \overline{L^c} = \Phi^i\varepsilon_{ij} \overline{L^{cj}}$, with $\varepsilon_{12}=1$. While the first two terms in Eq.~\eqref{eq:lag5_ewsb} signify double- and single-Higgs couplings to neutrinos of flavors $\ell\ell'$, the third term  generates the  $3\times3$ LH  Majorana mass matrix
$m_{\ell\ell'}$, as defined in Eq.~\eqref{eq:dim5Mass}.
After rotating $m_{\ell\ell'}$ into the mass basis, the resulting eigenvalues parametrize the three neutrino mass eigenstates $m_{\nu_k}$ that describe neutrino oscillation data.

We make no assumption on the structure of $C_5^{\ell\ell'}$. It is therefore possible under this framework that one neutrino is massless, as allowed by data~\cite{Wyler:1982dd}; that all masses scale as $m_{\nu_k}\sim\mathcal{O}(m_{\ell\ell'})$, indicating minor fine tuning;  or that $m_{\nu_k}\ll m_{\ell\ell'}$, indicating strong cancellations among the $m_{\ell\ell'}$ elements. As nuclear searches are only sensitive to $\vert m_{ee}\vert$, the latter possibilities remain under-explored.\\


\textbf{\textit{The $\nuBB$ Process at Dimension Five}}~-- A goal of this work is to estimate the sensitivity of the Large Hadron Collider (LHC) to the $\nuBB$ process, and hence the Weinberg operator.  When simulating the Weinberg operator at the LHC, difficulties arise if working in the neutrinos' mass eigenbasis. There, $d=5$ vertices  are proportional to $m_{\nu_k}$, which are unknown and small on  LHC  scales, and to $U_{\ell k}$, which carry unknown phases. So while the transition in Fig.~\ref{fig:wwScattWeinberg5_diagram}  may proceed through a non-trivial incoherent sum of intermediate states, individual contributions may be too small for practical computations. 

We propose a solution to this complication by working in the neutrino flavor basis and treating the mass term in  Eq.~\eqref{eq:lag5_ewsb} as a ``two-point vertex''. From this perspective, the  Weinberg operator in Fig.~\ref{fig:wwScattWeinberg5_diagram} couples one massless, LH neutrino of momentum $p$ and flavor $\ell$ with the conjugate of a second neutrino of momentum $p$ and flavor $\ell'$. After contracting Dirac matrices, the LN-violating $(\nu_\ell \nu_{\ell'}^c)$  current in Fig.~\ref{fig:wwScattWeinberg5_diagram} reduces to the ratio of  $m_{\ell\ell'}$ and the squared virtuality $p^2$.  Explicitly, its graph simplifies to:\\
\includegraphics[width=\columnwidth]{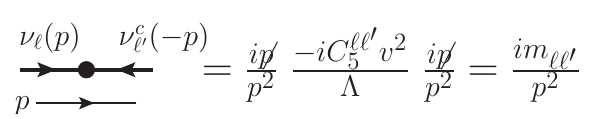}

Up to corrections of $\mathcal{O}(\vert m_{\ell\ell'}^2/p^2\vert)$, which are assumed small, one can identify the rightmost ratio as the right-handed (RH) helicity state of an intermediate fermion with mass $m_{\ell\ell'}$ and momentum $p$. That is, one can write
\begin{align}
\gamma^\alpha P_L\frac{ i\left(\not\!p +  m_{\ell\ell'}\right)}{p^2 - m_{\ell\ell'}^2} \gamma^\beta P_R & = \gamma^\alpha P_L \frac{ i m_{\ell\ell'}}{p^2 - m_{\ell\ell'}^2}P_L \gamma^\beta 
\label{eq:propFull}\\
= \gamma^\alpha P_L \frac{ i m_{\ell\ell'}}{p^2}P_L \gamma^\beta & \times\left[1 + \mathcal{O}\left(\left\vert\frac{m_{\ell\ell'}^2}{p^2}\right\vert\right)\right],
\label{eq:propExpansion}
\end{align}
where $P_{R/L}=\frac{1}{2}(1\pm\gamma^5)$ are the usual chiral projection operators in four-component notation, and recover the same ratio at leading power of the expansion. Intuitively, this identification follows from the inversion of helicity in LN-violating currents as discussed in Refs.~\cite{Kayser:1982br,Mohapatra:1998rq,Denner:1992vza,Denner:1992me,Han:2012vk,Ruiz:2020cjx,Fuks:2020att}.

As a result, up to corrections of  $\mathcal{O}(\vert m_{\ell\ell'}^2/p^2\vert)$,  the $(\nu_\ell \nu_{\ell'}^c)$ current itself  can be modeled as an unphysical Majorana neutrino $N$ with mass $m_{\ell \ell'}$  that couples to the $W$ boson and all charged leptons $\ell$ via the Lagrangian
\begin{align}
\Delta \mathcal{L} =
-\frac{g_W}{\sqrt{2}} W^+_\mu \sum_{\ell = e}^{\tau} \overline{N} \gamma^\mu P_L \ell^- + {\rm H.c.} 
\label{eq:lagCC}
\end{align}
Here, $g_W\approx0.65$ is the SU$(2)_L$ weak coupling constant. Up to factors of active-sterile mixing, Eq.~\eqref{eq:lagCC} is identical to the interaction Lagrangian in the Phenomenological Type I Seesaw model~\cite{delAguila:2008cj,Atre:2009rg}, and therefore can also be employed in LN-violating decays of hadrons and  leptons.\\

\textbf{\textit{Signal and Background Simulation}}~--
To simulate the $\nuBB$ process in LHC collisions using mainstream MC tools, we exploit the above observation that the intermediate $(\nu_\ell \nu_{\ell'}^c)$ current in Fig.~\ref{fig:wwScattWeinberg5_diagram} can be modeled as an unphysical Majorana neutrino with mass  $m_{\ell \ell'}=C^{\ell\ell'}_5 v^2/\Lambda$. We implement the Lagrangian of Eq.~\eqref{eq:lagFull} into the \fr~software package (version 2.3.36)~\cite{Christensen:2008py,Christensen:2009jx,Degrande:2011ua,Alloul:2013bka} by extending the  \fr~implementation of the SM (version 1.4.7) by a single Majorana neutrino $N$ with mass $m_N$ and EW boson couplings governed by Eq.~\eqref{eq:lagCC}. We ensure that conventional factors are kept according to Ref.~\cite{Alloul:2013bka}. To account for all  $\ell^\pm \ell^{'\pm}$ flavor permutations accessible at LHC energies, we make $m_N$ an internally calculated quantity that is set by
{
\begin{equation}
m_N = \left\vert C^{ee}_5+C^{e\mu}_5+C^{e\tau}_5+C^{\mu\mu}_5+C^{\mu\tau}_5+C^{\tau\tau}_5\right\vert\frac{v^2}{\Lambda}.
\label{eq:heavyNMass}
\end{equation}
}

Using Refs.~\cite{Degrande:2014vpa,Hahn:2000kx}, we extract renormalization and $R_2$  counterterms  up to the first order in the quantum chromodynamic (QCD) coupling $\alpha_s$. Feynman rules are collected into a set of public universal \fr~ output (UFO) libraries that we call the \libName~libraries. 

With this UFO proton collisions are simulated at next-to-leading (NLO) in QCD with the event generator \textsc{\small MadGraph5\_aMC@NLO}~(version 2.7.1.2)~\cite{Stelzer:1994ta,Alwall:2014hca,Frixione:2002ik,Frederix:2009yq,Hirschi:2011pa,Hirschi:2015iia}. Parton showering (PS) and  modeling of non-perturbative phenomena are handled by \textsc{Pythia8} (version 243)~\cite{Sjostrand:2014zea}. Hadron-level events are passed through \textsc{Delphes} (version 3.4.2)~\cite{deFavereau:2013fsa} for the simulation of an ATLAS-like detector. Hadron clustering is handled according to the anti-$k_T$ algorithm at $R=0.4$~\cite{Catani:1993hr,Ellis:1993tq,Cacciari:2008gp} as implemented in \textsc{FastJet}~\cite{Cacciari:2005hq,Cacciari:2011ma}. We tune our simulation tool chain as in the study on $W^\pm W^\pm$ scattering by Ref.~\cite{Fuks:2020att}, whose methodology we also follow to model SM backgrounds.\\


\textbf{\textit{The $d=5$, $\nuBB$ Process at the LHC}}~-- In LHC collisions the LN-violating $\nuBB$ process occurs  through the scattering of two same-sign $W$ bosons that are sourced from quarks and antiquarks, and exit as two high-$p_T$ jets.  At the hadronic level, the collider signature is given by
\begin{equation}
p ~p ~\to ~j ~j ~\ell^\pm  ~\ell^{'\pm} +X,
\label{eq:nuBBIncl}
\end{equation}
where $X$ represents the additional hadronic and electromagnetic activity that may exist in the inclusive process. 

To identify the  dependence of Eq.~\eqref{eq:nuBBIncl} on the Weinberg operator, we consider the Effective $W$ Approximation~\cite{Dawson:1984gx,Kane:1984bb, Kunszt:1987tk} and treat the incoming $W^\pm W^\pm$ pair as perturbative constituents of the proton. In this limit, we find that the $W^\pm W^\pm \to \ell^\pm \ell^{'\pm}$ sub-process is dominated by the scattering of longitudinal $W$ bosons. After summing over all external helicities, the spin-averaged, parton-level cross section for the $2\to2$ process is given by
\begin{align}
\hat{\sigma}(W^+ W^+ &\to \ell^+ \ell^{'+}) 
\nonumber\\
&= \frac{(2-\delta_{\ell\ell'})}{2\pi ~ 3^2}\left\vert\frac{C_5^{\ell\ell'}}{\Lambda}\right\vert^2
+ \mathcal{O}\left(\frac{m_{W}^2}{M_{WW}^2}\right).
\label{eq:wwScattXSec}
\end{align}
This shows that like in nuclear experiments  the $\nuBB$ rate at the LHC scales as $\sigma\sim\vert m_{\ell\ell'}\vert ^2 \propto \vert C^{\ell\ell'}_5/\Lambda\vert^2$.

Using this scaling behavior, we have checked that setting $\Lambda\ll200\TeV$ in simulations with the \libName~UFO will generate unphysical cross sections. This is due to a breakdown of the expansion in Eq.~\eqref{eq:propExpansion},
which requires $v^2/\Lambda$ to be small compared to the  virtuality of
the internal $(\nu_\ell \nu_{\ell'}^c)$  current. For the LHC and beyond, physical rates can be obtained by choosing, for example, $\Lambda=200\TeV$ and using the relationship
\begin{equation}
    \sigma(\Lambda) = \sigma(\Lambda=200\TeV) \times \left(\frac{200\TeV}{\Lambda}\right)^2.
\end{equation}

\begin{figure}[!t]
\includegraphics[width=\columnwidth]{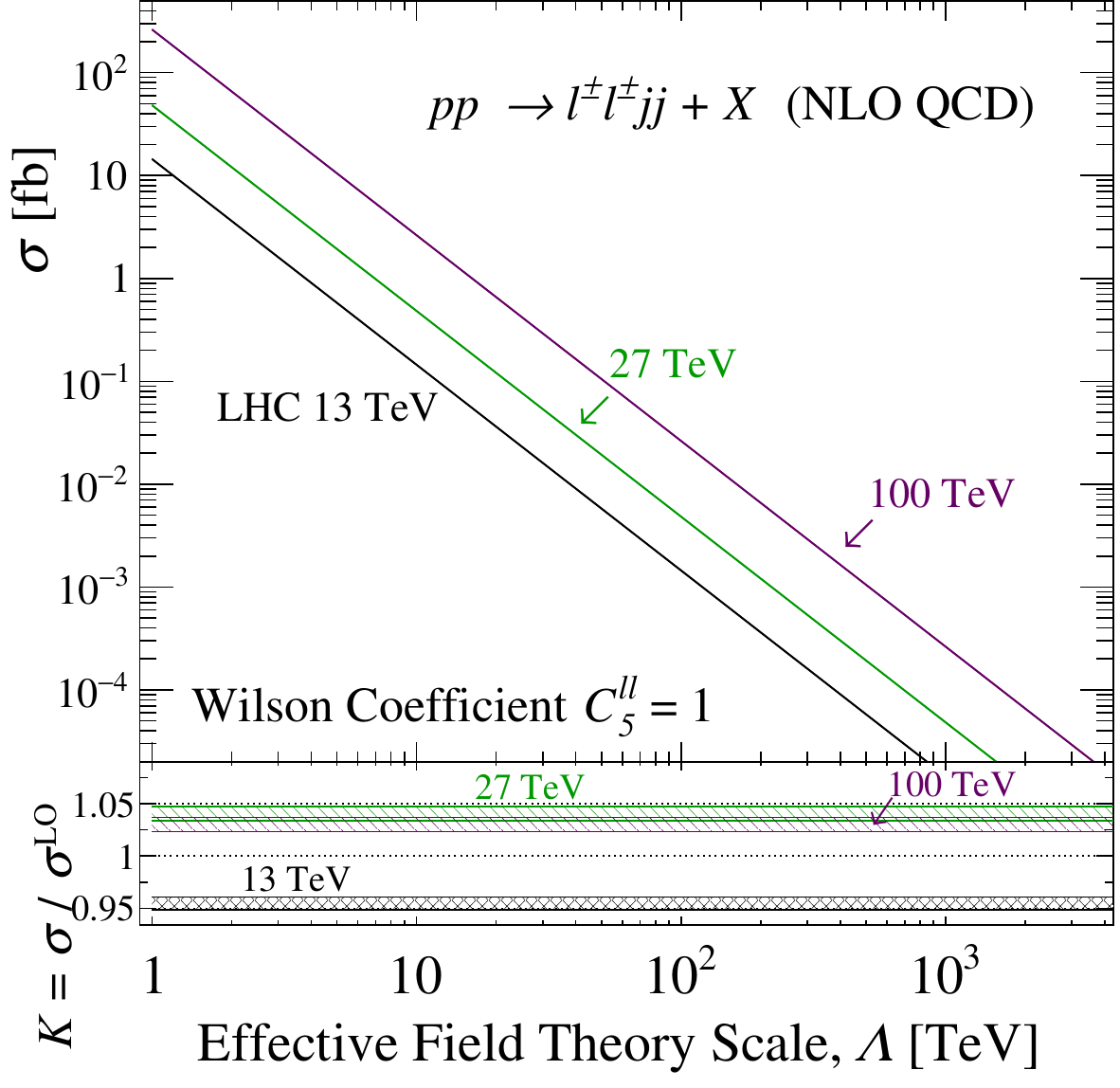}
\caption{
  Total cross sections at NLO in QCD (top) and the corresponding NLO $K$-factors (bottom) for the process in Eq.~\eqref{eq:nuBBIncl}, as a function of EFT scale $\Lambda$ with $C^{\ell\ell'}_5=\delta_{\ell\mu}\delta_{\ell'\mu}$, and  $\sqrt{s} = 13$, 27, 100~TeV. Bands represent scale uncertainties.}
\label{fig:wwScattD5_XSec_vs_EFT_LHCMulti}
\end{figure}

Given this guidance, we show in the top of Fig.~\ref{fig:wwScattD5_XSec_vs_EFT_LHCMulti} the hadronic cross section $\sigma$ at NLO in QCD for the full $2\to4$, $\nuBB$ process for $\sqrt{s}=13,~27$ and 100~TeV, as a function of $\Lambda$, assuming $C_5^{\ell\ell'} = \delta_{\ell\mu}\delta_{\ell'\mu}$. The band thickness for each curve, which reaches $\mathcal{O}(0.5\%-1.5\%)$, denotes the nine-point scale uncertainty. While not shown, PDF uncertainties reach about $\mathcal{O}(1\%)$. At $\sqrt{s}=13\TeV$ and for $\Lambda=10\TeV$ (or $m_{\mu\mu}\sim6\GeV$) we find $\sigma\sim0.14\fb$. Conversely, at $\sqrt{s}=13~(27)~[100]\TeV$ we find the rate  reaches the $\sigma\sim1\ab$ threshold at $\Lambda\sim120~(220)~[510]\TeV$, which corresponds to $m_{\mu\mu}\sim500~(275)~[120]\MeV$. As  a measure of the   QCD corrections to the cross section, we show in the bottom of Fig.~\ref{fig:wwScattD5_XSec_vs_EFT_LHCMulti}  the  QCD $K$-factor,  defined as the ratio of the NLO and leading order (LO) cross sections. We report that $\mathcal{O}(\alpha_s)$ corrections are mild, reaching $K\sim 0.95-1.05$ across $\sqrt{s}$.

To estimate the LHC's discovery potential of the Weinberg operator, we focus on the $\ell\ell'=\mu\mu$ channel with benchmark inputs $C_5^{\ell\ell'} = \delta_{\ell\mu}\delta_{\ell'\mu}$ and $\Lambda=200\TeV$ (or $m_{\mu\mu}\approx300\MeV$), and design an analysis inspired by the Run 2 performance of the ATLAS detector~\cite{PERF-2007-01,ATLAS-TDR-2010-19}. We employ particle identification requirements on electrons, muons, and jets that are summarized in the top  of Table~\ref{tab:cut_summary}. For simplicity, we ignore particles originating from pileup interactions as they would mostly be subtracted with dedicated algorithms in real experiments.

\begin{figure}[!t]
  \centering
  \includegraphics[width=\columnwidth]{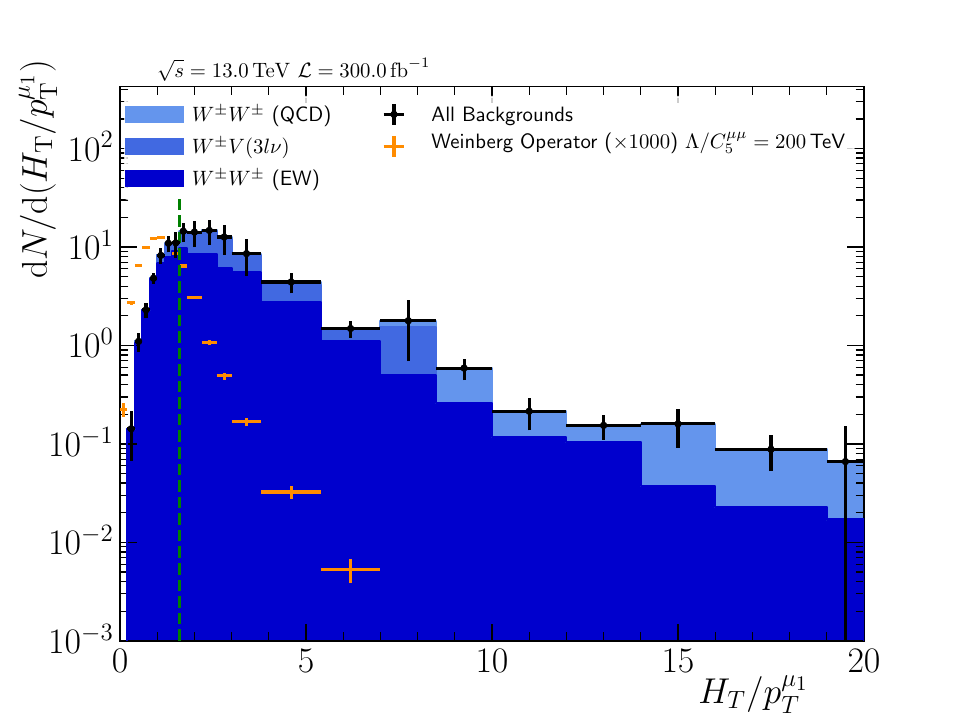}
    \caption{At $\sqrt{s}=13\TeV$  with $\mathcal{L}=300\invfb$, the  $(H_T/p_T^{\mu_1})$ distribution at NLO+PS for signal and backgrounds in the signal region, for $C_5^{\ell\ell'} = \delta_{\ell\mu}\delta_{\ell'\mu}$ and $\Lambda=200\TeV$.}
    \label{fig:kinematics_preselection}
\end{figure}

To define our signal-enriched region we demand events to have at least two jets, with the leading pair carrying a large invariant mass, and exactly two muons with the same charge. Events with additional leptons, including hadronically decaying $\tau$ leptons, are vetoed. To further reduce  backgrounds we take into account two qualitative differences between our signal and background processes:
(i) Unlike SM processes with the same final state, our signal does not contain outgoing neutrinos. As neutrinos go undetected in LHC experiments, their presence give rise to missing transverse momentum $\etmiss$, which is defined as the $p_T$ recoil against all visible objects.  We therefore require that events have a small $\etmiss$, in accordance with the detector resolution.
(ii) Due to the lack  of QCD color flowing between the two hadrons in Fig.~\ref{fig:wwScattWeinberg5_diagram}, the hadronic activity  is much milder than  the QCD and $W^\pm V$ backgrounds. Following past studies~\cite{Pascoli:2018rsg,Pascoli:2018heg,Fuks:2019iaj}, we impose an upper limit on the ratio $(H_T/p_T^{\mu_1})$, where $H_T$ is the scalar sum of jet $p_T$. To guide our precise cut choice, we plot in Fig.~\ref{fig:kinematics_preselection} the $(H_T/p_T^{\mu_1})$ distribution for our signal and leading backgrounds after applying all other selection cuts.

\begin{table}[!t]
  \renewcommand{\arraystretch}{1.1}
    \caption{Particle identification and signal region definitions}
    \centering
    \resizebox{\columnwidth}{!}{
    \begin{tabular}{c}
    \hline  \hline
    Particle Identification Cuts\\
    $p_T^{e~(\mu)~[j]} > 10~(10)~[25]\GeV$, \quad
    Anti-$k_T$($R$=0.4)\\
    $\vert\eta^{e~(\mu)~[j]}\vert<2.5~(2.7)~[4.5]$ \\
	\hline
    Signal Region Cuts \\ 
    $p_T^{\mu_1~(\mu_2)} > 27~(10)\GeV$,     ~ 
    $n_\mu=2$,  ~
    $n_j\geq2$, \\ $n_e = n_{\tau^\mathrm{had}} =0$, ~
    $Q_{\mu_1}\times Q_{\mu_2}=1$, ~  
    $M(j_1,j_2) > 700 \GeV$\\
    $\etmiss < 30\GeV$, $(H_\mathrm{T}/p_\mathrm{T}^{\mu_1}) < 1.6$ \\
    \hline
    Changes to Identification and Signal Cuts at $\sqrt{s}=100\TeV$\\ 
    $\vert\eta^{e~(\mu)~[j]}\vert<4.0~(4.0)~[5.5]$,  $M(j_1,j_2) > 1\TeV$\\
    $\etmiss < 20\GeV$, $(H_\mathrm{T}/p_\mathrm{T}^{\mu_1}) < 0.6$\\
    \hline  \hline
    \end{tabular}
    }
    \label{tab:cut_summary}
\end{table}

At this stage, the leading backgrounds consist of mixed EW-QCD production of $W^\pm W^\pm jj$, pure EW production of $W^\pm W^\pm jj$,  and the inclusive diboson+jets spectrum $W^\pm V + nj$, with $V\in\{\gamma^*/Z/Z^*\}$. We checked that other processes, {\it e.g.}, $t\overline{t}W^\pm$, do not appreciably survive our event selection. We neglect processes that are especially difficult to simulate from MC methods alone. This includes when muons are assigned the wrong charge during event reconstruction. While sub-dominant for dimuon final states, such backgrounds are  relevant for the electron and tau channels~\cite{Aad:2016tuk,Alvarez:2016nrz,CMS-DP-2017-036,Pascoli:2018heg}. We account for such backgrounds with a more conservative uncertainty in our background estimate. We summarize our signal region definition in Table~\ref{tab:cut_summary}. About $\varepsilon\sim12\%$ of generated signal events with $\Lambda=200\TeV$ pass all identification and signal region cuts. For our inputs the signal (total background) rate reaches $\sigma\sim42$~zb~$(\sigma\sim25\ab)$.\\

\textbf{\textit{Sensitivity to the Weinberg Operator}}~-- 
To quantify any excess of events, we apply a Poisson-counting likelihood with a background rate uncertainty that is constrained by an auxiliary Poisson measurement~\cite{ATLAS:2020yaz, Cousins:2008zz}.  Assuming a $\delta_b = 20\%$ systematic uncertainty in the background, we estimate the sensitivity  at 95\% CL  to $\vert C_5^{\mu\mu}\vert/\Lambda\propto m_{\mu\mu}$ by fixing our signal significance to $Z\approx2$ and the number of signal events $n_s$ to
\begin{equation}
    n_s = n_s^0 \times \vert C_5^{\mu\mu}\vert ^2\left(\frac{200\TeV}{\Lambda}\right)^2,
\end{equation}
where $n_s^0$ is the number of signal events for our benchmark inputs. We then solve this equality for $\vert C_5^{\mu\mu}\vert$. With $\mathcal{L}=300\invfb~(3\invab)$, we report that the LHC (HL-LHC) is sensitive at 95\% CL to scales below
\begin{equation}
\confirm{\Lambda / \vert C_5^{\mu\mu}\vert  \lesssim 8.3~(11)\TeV.}
\label{eq:lhcSensitivity}
\end{equation}
These translate into effective $\mu\mu$ Majorana masses of \begin{equation}
\confirm{
\vert m_{\mu\mu}\vert \gtrsim 7.3~(5.4) \GeV.
}
\end{equation}

With an outlook to potential successors of the HL-LHC~\cite{Strategy:2019vxc,EuropeanStrategyGroup:2020pow}, we estimate the sensitivity of a $\sqrt{s}=100\TeV$ proton collider. We employ our LHC analysis but with changes listed at the bottom of Table~\ref{tab:cut_summary}. We set $\delta_b = 5\%$ to account for improved detector resolution and control region modeling. For $\mathcal{L}=30\invab$ of data, we find sensitivity to \confirm{$\Lambda / \vert C_5^{\mu\mu}\vert  \lesssim 48\TeV$} at 95\% CL. Our precise choice of cuts are for illustration and optimization should be investigated. This is especially relevant as we neglect an $\mathcal{O}(30\%)$ statistical uncertainty on our $WV^\pm$ simulation despite starting from \confirm{$10^7$} NLO+PS events.

As described above, treating the Weinberg operator as an unphysical Majorana fermion is applicable to LN-violating decays of mesons, so long as the expansion in Eq.~\eqref{eq:propExpansion} is satisfied. 
Using Refs.~\cite{Atre:2005eb,Bediaga:2018lhg,CortinaGil:2019dnd}, we update the limits and projections on $\vert m_{\mu\mu}\vert$ from $B^\pm \to \pi^\pm \mu^\mp\mu^\mp$ and $K^\pm\to \pi^\mp\mu^\pm\mu^\pm$ decays. We find that LHCb with $\mathcal{L}=300\invfb$ can only probe $\Lambda/\vert C_5^{\mu\mu}\vert\lesssim 9\MeV$ while NA-62 has excluded with its 2017 data set $\Lambda/\vert C_5^{\mu\mu}\vert\lesssim 1.1\TeV$. 

\begin{figure}[!t]
    \centering
    \includegraphics[width=\columnwidth]{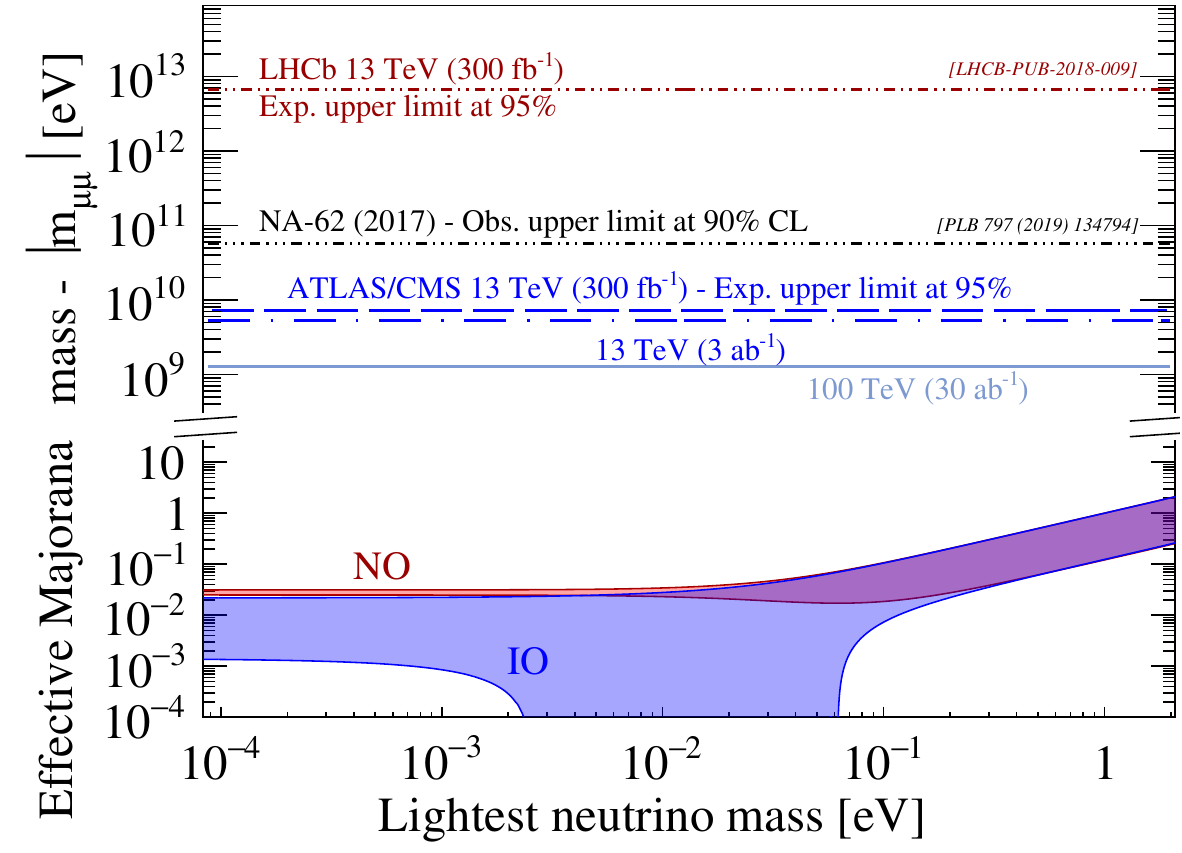}
    \caption{
    Projected sensitivity to $\vert m_{\mu\mu}\vert$ at the $\sqrt{s}=13\TeV$ LHC and a successor proton collider at $\sqrt{s}=100\TeV$, observed limits set by NA-62 with its 2017 data set~\cite{CortinaGil:2019dnd}, and allowed values by best-fits to neutrino oscillation data~\cite{Esteban:2020cvm}. 
      }
    \label{fig:wwScattD5_discovery}
\end{figure}

Assuming that neutrino masses are described completely at $d=5$, we summarize in Fig.~\ref{fig:wwScattD5_discovery} our sensitivities to $\vert m_{\mu\mu}\vert$ in comparison to the values~\cite{Esteban:2020cvm} allowed by Eq.~\eqref{eq:effMajMass} (generalized for arbitrary $\ell\ell'$~\cite{Frigerio:2002rd}) for normal (NO) and inverse ordering (IO) of neutrino masses. The reach of $W^\pm W^\pm$ scattering greatly exceeds our LHCb and NA-62 benchmarks. Nevertheless, improvements at these and similar experiments are anticipated.\\


\textbf{\textit{Conclusions}}~-- If the Weinberg operator is present in nature and is accessible at collider energies, then a key prediction are processes that violate LN, such as the $\nuBB$ transition, and possibly charged lepton flavor. Motivated by the flavor limitations of nuclear decay experiments,  we have investigated the LHC's sensitivity to the Weinberg operator using the $W^\pm W^\pm \to \ell^\pm_i\ell^\pm_j$ process, which permits muon- and tau-flavored final states. We find  sensitivity that exceeds representative searches at $B$- and $K$-meson factories, and establishes a complementarity across accelerator facilities in the search for the Weinberg operator.\\


\textbf{\textit{Acknowledgements}}~--  
JN and KP acknowledge support by the Deutsche Forschungsgemeinschaft (DFG, German Research Foundation) under Germany's Excellence Strategy -- EXC 2121 ``Quantum Universe'' – 390833306.
RR acknowledges the support of Narodowe Centrum Nauki under Grant No. 2019/34/E/ST2/ 00186, and the UCLouvain fund ``MOVE-IN Louvain.'' 
This work  received funding from the European Union's Horizon 2020 research and innovation programme as part of the Marie Skłodowska-Curie Innovative Training Network MCnetITN3 (grant agreement No.~722104), FNRS “Excellence of Science” EOS be.h Project No.~30820817.  Computational resources have been provided by the supercomputing facilities of the Universit\'e catholique de Louvain (CISM/UCL)  and the Consortium des \'Equipements de Calcul Intensif en F\'ed\'eration Wallonie Bruxelles (C\'ECI) funded by the Fond de la Recherche Scientifique de Belgique (F.R.S.-FNRS) under convention 2.5020.11 and by the Walloon Region. \\

\appendix
\textbf{\textit{Appendix: Technical details on methodology}}~--  In this appendix we provide additional discussions and details of our methodology. 

In a generic gauge, the Higgs field in terms of the EW Goldstone bosons $G^{\pm,0}$ is $\sqrt{2}\Phi = (-i\sqrt{2}G^+, v+h+iG^0)^T$. Explicit contraction of SU$(2)_L$ indices then gives
\begin{align}
    L_{\ell}\!\!\cdot\!\Phi &= L_{\ell}^i\varepsilon_{ij}\Phi^j 
    = \frac{1}{\sqrt{2}}\nu_{\ell}(v+h+iG^0) + i \ell G^+, \\
    \Phi\!\cdot\! \overline{L}^c_{\ell } &= 
    -i G^+ \overline{\ell^c} -  \frac{1}{\sqrt{2}}(v+h+iG^0)\overline{\nu^c_\ell}.
\end{align}
This allows us to express the full Weinberg operator as
\begin{align}
 \lag_5 = 
    &-\frac{C_5^{\ell\ell'}}{2\Lambda} \left(v^2 + 2vh + hh\right)\overline{\nu^c_\ell}\nu_{\ell'} 
    \\
    &- \frac{i C_5^{\ell\ell'}}{\sqrt{2}\Lambda} G^+(v+h) (\overline{\nu^c_\ell}\ell'+\overline{\ell^c}\nu_{\ell'})
    \\
    &-\frac{iC_5^{\ell\ell'}}{\Lambda}G^0(v+h)\overline{\nu^c_\ell}\nu_{\ell'}    
    \\
    &+ \frac{C_5^{\ell\ell'}}{2\Lambda} \left(2G^+G^+\overline{\ell^c}\ell'  + G^0G^0\overline{\nu^c_\ell}\nu_{\ell'}
 \right)
    \\
    &+ \frac{C_5^{\ell\ell'}}{\sqrt{2}\Lambda} G^0G^+ (\overline{\nu^c_\ell}\ell'+\overline{\ell^c}\nu_{\ell'})
     + \text{H.c.}
\end{align}
Here and below, the Hermitian conjugate is understood to apply to the full expression, not simply the final line.

With this,  the interaction Lagrangian by which we extend the SM Lagrangian in the \libName~UFO is
\begin{align}
&\Delta \mathcal{L} =
-\frac{g_W}{\sqrt{2}} W^+_\mu \sum_{\ell = e}^{\tau} \overline{N} \gamma^\mu P_L \ell^- 
\\
&-\frac{g_W}{2\cos\theta_W} Z_\mu \sum_{\ell = e}^{\tau} \overline{N} \gamma^\mu P_L \nu_\ell 
\\
&-\frac{g_W m_N}{2m_W} h\left(1+\frac{g_W}{4m_W}h\right) \sum_{\ell = e}^{\tau} \overline{N}  P_L \nu_\ell 
\\
&-i\frac{g_W m_N}{2\sqrt{2}m_W} G^+\left(1+\frac{g_W}{2m_W}h\right) \sum_{\ell = e}^{\tau} \left(\overline{N}  P_L \ell+ \overline{\ell^c}P_L N\right)
\\
&-i\frac{g_W m_N}{2m_W} G^0\left(1+\frac{g_W}{2m_W}h\right)\sum_{\ell = e}^{\tau} \overline{N}  P_L \nu_\ell 
\\
&+\frac{g_W^2 m_N}{8m_W^2} \left( 2G^+G^+ \sum_{\ell,\ell' = e}^{\tau}\overline{\ell^c}P_L\ell'+ G^0G^0 \sum_{\ell=e}^{\tau}\overline{N}P_L\nu_{\ell}
\right)
\\
&+\frac{g_W^2 m_N}{4\sqrt{2}m_W^2} G^0G^+\sum_{\ell = e}^{\tau} \left(\overline{N}  P_L \ell+ \overline{\ell^c}P_L N\right) + {\rm H.c.}
\end{align}

To further understand the identification in Eq.~\eqref{eq:propExpansion}, we recall that the fermions in the LN-violating $(\ell^+ \nu_\ell \nu_{\ell'}^c \ell^{'+})$ current in Fig.~\ref{fig:wwScattWeinberg5_diagram} experience an additional parity inversion beyond the standard SU$(2)_L$ chiral couplings~\cite{Kayser:1982br,Mohapatra:1998rq}. In terms of Feynman rules~\cite{Denner:1992vza,Denner:1992me}, this manifests as a chiral inversion of the $(W\ell'\nu_{\ell'}^c)$ vertex, \ie, $\gamma^\beta P_L \to \gamma^\beta P_R$. In the absence of additional new physics, this ensures~\cite{Han:2012vk,Ruiz:2020cjx} the presence of the $P_{R/L}$ projection operators that envelope the $(\nu_\ell \nu_{\ell'}^c)$ current in Eq.~\eqref{eq:propFull}, and hence that the $(\nu_\ell \nu_{\ell'}^c)$ current propagates in the RH helicity state.

\begin{table}[!t]
\begin{center}
\caption{Parameters for the \libName~UFO. \label{tab:ufoParams}}
  \resizebox{\columnwidth}{!}{
\setlength\tabcolsep{4pt}
\begin{tabular}{c c c c c}
\hline\hline
  Parameter & FR name & Type  & LH block & LH counter\\
  \hline
  $\Lambda$   		& {\tt Lambda}	& External (Real)	& {\tt NUPHYSICS}	& 1\\
  $C_5^{\ell\ell'}$   	& {\tt Cll}		& External (Real)	& {\tt NUPHYSICS}	& 2-7\\
  $m_N$			& {\tt mN}		& Internal (Real)	& {\tt MASS}		& 9900012\\
  $\Gamma_N$    & {\tt wN}      & Internal (Real)   & {\tt WIDTH}       &
  9900012\\
 \hline \hline
\end{tabular}
}
\end{center}
\end{table}

The model's input parameters along with their \fr~and Les Houches~\cite{Skands:2003cj} information are summarized in Table~\ref{tab:ufoParams}. The syntax used to import the UFO into \textsc{\small MadGraph5\_aMC@NLO}~and simulate  the process
\begin{eqnarray}
q_1 ~q_2 ~ \to ~ q'_1 ~q'_2 ~\mu^\pm ~\mu^\pm,
\end{eqnarray}
where $q$ is any light quark or antiquark, at NLO is
\begin{verbatim}
import model SMWeinbergNLO
generate     p p > mu+ mu+ j j QED=4 QCD=0
    $$ w+ w- [QCD]
add process  p p > mu- mu- j j QED=4 QCD=0 
    $$ w+ w- [QCD]
\end{verbatim}

For SM inputs, we approximate the quark sector by $n_f=5$ massless quarks that do not mix. 
Values of couplings and masses are set to global averages reported in the 2020 Particle Data Group review~\cite{Zyla:2020zbs}:
\begin{align}
m_t(m_t) &= 172.76\GeV,    m_h = 125.1\GeV,  \label{eq:sminputs}\\
M_Z 	&= 91.1876\GeV,   \alpha_{\rm QED}^{-1}(M_Z)=127.952, \nonumber\\
G_F     &= 1.1663787\cdot 10^{-5}\GeV^{-2}, ~ \alpha_s(M_Z)=0.118.\nonumber
\end{align}

We employ the NNPDF3.1 NLO+LUXqed parton distribution function set  \texttt{(lhaid=324900)} \cite{Manohar:2016nzj,Manohar:2017eqh,Bertone:2017bme}, with scale evolution driven by \textsc{LHAPDF}~\cite{Buckley:2014ana}, and
PDF uncertainties are extracted using the replica method~\cite{Buckley:2014ana,Bertone:2017bme}. We fix the collinear factorization $(\mu_f)$, QCD renormalization $(\mu_r)$, and shower matching $(\mu_s)$ scales to the default values in Ref.~\cite{Alwall:2014hca}. The uncertainty in choosing $\mu_f$ and $\mu_r$ is quantified by scaling their baseline values by factors of 0.5, 1 and 2 to obtain a nine-point uncertainty band.

\begin{table}[!t]
\begin{center}
 \scriptsize
\caption{
The total cross section at NLO in QCD for the process in Eq.~\eqref{eq:nuBBIncl} for various choices of EFT scale $\Lambda$ and collider energy $\sqrt{s}$ at a Wilson coefficient $C_5^{\ell\ell'} = \delta_{\ell\mu}\delta_{\ell'\mu}$, and the corresponding mass $m_N$, uncertainties, and NLO $K$-factor.  
}\resizebox{\columnwidth}{!}{
\begin{tabular}{c c c c c c c}
\hline\hline
$\Lambda$ [TeV]  & $m_N$ [GeV] 	& $\sqrt{s}$ [TeV]
& $\sigma^{\rm NLO}$ [ab] & $\delta_{\rm Scale}$ 
& $\delta_{\rm PDF}$ 
& $K^{\rm NLO}$ \\
\hline
10	& 6		& 13		& $133$		& $^{+0.8\%}_{-0.8\%}$ & $^{+0.9\%}_{-0.9\%}$   & $0.968$ \\
100	& 0.6		& 13		& $1.42$		& $^{+1.0\%}_{-0.6\%}$ & $^{+0.9\%}_{-0.9\%}$   & $0.978$ \\
200 	& 0.3		& 13		& $0.361$   	& $^{+0.7\%}_{-0.6\%}$ & $^{+1.0\%}_{-1.0\%}$   & $0.952$ \\
400 	& 0.15 	& 13 		& $0.0904$  	& $^{+0.6\%}_{-0.8\%}$ & $^{+0.9\%}_{-0.9\%}$   & $0.988$ \\
200 	& 0.3 	& 27		& $1.21$    & $^{+0.9\%}_{-0.8\%}$ & $^{+0.9\%}_{-0.9\%}$   & $1.04$ \\
200	& 0.3  	& 100	& $6.56$    & $^{+1.4\%}_{-1.2\%}$ & $^{+0.9\%}_{-0.9\%}$   & $1.03$ \\
\hline\hline
\end{tabular}
} 
 \label{tab:eftxsec}
\end{center}
\end{table}

As a check of the \libName~UFO, we consider the amplitudes for  the  $W^\pm(p^W_1,\lambda^W_1) W^\pm(p^W_2,\lambda^W_2) \to \ell^\pm(p_1^\ell,\lambda^\ell_1)  \ell^{'\pm}(p_2^\ell,\lambda^\ell_1)$ process. Explicit calculation reveals that in the high-energy limit, \ie,~ when $M_{WW}^2=(p_1^W+p_2^W)^2\gg m_W^2$, the $2\to2$ process is dominated by the scattering of two longitudinally polarized $W^\pm$ bosons. 
For the $(\lambda_1^W,\lambda^W_2)=(0,0)$ helicity  configuration, the exact helicty amplitude is
\begin{align}
-i\mathcal{M}&(W_0^+W_0^+\to \ell^{+}_R\ell^{'+}_R) =
-i\mathcal{M}_t + -i\mathcal{M}_u,
\\
-i\mathcal{M}_t& = ie^{-i\phi_1}\left(\frac{C_5^{\ell\ell'} }{\Lambda}\right)\left(\frac{M_{WW}^3}{ t}\right)
\nonumber\\
& \quad \times \left[1-2r_W-\sqrt{1-4r_W}\cos\theta_1\right],
\\
-i\mathcal{M}_u& = ie^{-i\phi_1}\left(\frac{C_5^{\ell\ell'} }{\Lambda}\right)\left(\frac{M_{WW}^3}{ t}\right)
\nonumber\\
& \quad \times \left[1-2r_W+\sqrt{1-4r_W}\cos\theta_1\right],
\end{align}
where $r_W = m_W^2 / M_{WW}^2$; $\theta_1$ and $\phi_1$ are respectively the polar and azimuthal angles of $\ell(p_1^\ell)$ in the $(WW)$-frame, and the kinematic invariants are defined by $t = (p_1^W - p_1^\ell)^2$ and  $u = (p_1^W - p_2^\ell)^2$. Further evaluation of $t$ and $u$ results in the somewhat simple expression
\begin{equation}
    \mathcal{M}(W_0^+W_0^+\to \ell^{+}_R\ell^{'+}_R) = 
    e^{-i\phi_1}\left(\frac{4C_5^{\ell\ell'} M_{WW} }{\Lambda}\right).
\end{equation}
The $J=0$ partial wave is subsequently given by
\begin{align}
    a_{J=0} &= \frac{1}{32\pi}\int_{-1}^{1} ~d\cos\theta_1 ~ \mathcal{M}(W_0^+W_0^+\to \ell^{+}_R\ell^{'+}_R)
    \\
    &= \frac{1}{4\pi}\frac{C_5^{\ell\ell'}  M_{WW}}{\Lambda}.
\end{align}
Since $s$-wave perturbative unitarity requires that $\vert a_J\vert < 1$, one obtains the constraint that
\begin{equation}
\vert C_5 \vert M_{WW} < 4\pi \Lambda.
\end{equation}

After evaluating the exact helicity amplitude for each $(\lambda_1^W,\lambda^W_2)$ permutation, taking their sum, and then taking the high-energy limit, we obtain
\begin{align}
    \sum_{\{\lambda^W,\lambda^\ell\}}\left\vert\mathcal{M}(W^+W^+\to \ell^{+}\ell^{'+})\right\vert^2 &= 
    \nonumber\\
    8(2-\delta_{\ell\ell'})
    \left\vert \frac{C_5^{\ell\ell'} M_{WW}}{\Lambda} \right\vert^2 +& \mathcal{O}\left(\frac{m_W^2}{M_{WW}^2}\right).
\end{align}
The Kronecker $\delta$ accounts for the $1/2!$ symmetry factor needed for amplitudes with identical final-state particles. This implies a totally differential cross section of
\begin{equation}
    \frac{d\hat{\sigma}}{d\cos\theta_1 d\phi_1}  
    = 
    \frac{(2-\delta_{\ell\ell'})}{8\pi^2 3^2}
    \left\vert \frac{C_5^{\ell\ell'}}{\Lambda} \right\vert^2 + \mathcal{O}\left(\frac{m_W^2}{M_{WW}^2}\right).
\end{equation}
Integration over the full solid angle recovers Eq.~\eqref{eq:wwScattXSec}.

Using Eq.~\eqref{eq:wwScattXSec} as a check of the \libName~UFO, we list in Table~\ref{tab:eftxsec}  the total $2\to4$, hadronic cross section $\sigma$ for Eq.~\eqref{eq:nuBBIncl} at NLO in QCD  for representative cutoff scales $\Lambda$, assuming a Wilson coefficient of $C_5^{\ell\ell'} = \delta_{\ell\mu}\delta_{\ell'\mu}$, for the $\sqrt{s}=13\TeV$ LHC and proposed  experiments at $\sqrt{s}=27\TeV$ and 100\TeV. Also listed are the corresponding (unphysical) Majorana neutrino mass $m_N$, as defined by Eq.~\eqref{eq:heavyNMass}, the nine-point scale uncertainty  $\delta_{\rm Scale}$, the parton distribution function (PDF) uncertainty $\delta_{\rm PDF}$, and the QCD  $K$-factor, which is defined as the ratio $K = \sigma / \sigma^{LO}$,
where $\sigma^{\rm LO}$ is the LO rate.

For $\Lambda=10\TeV$ and $100\TeV$ at $\sqrt{s}=13\TeV$, we observe a cross section scaling of $\sigma(\Lambda=10\TeV)/\sigma(\Lambda=100\TeV)\sim93$, undershooting the $100\times$ scaling expected from Eq.~\eqref{eq:wwScattXSec}. We attribute this to a breakdown of Eq.~\eqref{eq:propExpansion}, which requires the mass $m_N\sim v^2/\Lambda$ to be small compared to the virtuality of $(\nu_\ell \nu_{\ell'}^c)$. At larger  $\Lambda$ we find, for example, that $\sigma(\Lambda=100\TeV)/\sigma(\Lambda=200\TeV)\sim3.93$ and $\sigma(\Lambda=200\TeV)/\sigma(\Lambda=400\TeV)\sim3.99$, indicating behavior more inline with Eq.~\eqref{eq:wwScattXSec}. We conclude that choices of $\Lambda\gtrsim200\TeV$ generate sufficiently small $m_N$ so that Eq.~\eqref{eq:propExpansion} remains valid for $\sqrt{s}\gtrsim 13\TeV$.

Assuming benchmark signal inputs of $\Lambda=200\TeV$ and $C_5^{\ell\ell'} = \delta_{\ell\mu}\delta_{\ell'\mu}$, we summarize in Table~\ref{tab:totalRates} the expected number of signal and background events after all cuts for the LHC (HL-LHC) with $\mathcal{L}=300\invfb~(3\invab)$. To quantify the LHC's sensitivity to the Weinberg operator, we define our signal significance $Z$ as~\cite{ATLAS:2020yaz, Cousins:2008zz}
\begin{align}
Z &= \frac{(n - n_b)}{\vert n - n_b\vert}  \sqrt{2\left[n\log x - \frac{n_b^2}{\delta_b^2}\log y\right]}, \quad \mathrm{with}
\label{eq:significance}
\\
x  &= \frac{n(n_b+\delta_b^2)}{n_b^2+n \delta_b^2}, \quad \mathrm{and} \quad
y  = 1+ \frac{\delta_b^2(n-n_b)}{n_b(n_b + \delta_b^2)}.
\end{align}
Here, $n = n_s + n_b$ is the total number of observed events, $n_s~(n_b)$ is the predicted number of signal (background) events, and $\delta_b$ is the uncertainty on $n_b$.

Under the parametrization of the PMNS matrix
\begin{align}
    U_{\rm PMNS} = \begin{pmatrix}
1 & 0 & 0\\ 
0 & c_{23} & s_{23}\\ 
0 & -s_{23} & c_{23}
\end{pmatrix}
&\cdot 
\begin{pmatrix}
c_{13}&0&s_{13}e^{-i\delta_{\rm CP}} \\
0&1&0 \\
-s_{13}e^{i\delta_{\rm CP}}&0&c_{13} \\
\end{pmatrix}
\nonumber\\
\cdot
\begin{pmatrix}
c_{12}&s_{12}&0 \\
-s_{12}&c_{12}&0 \\
0&0&1 \\
\end{pmatrix}
&\cdot
\begin{pmatrix}
e^{i\eta_1}&0&0 \\
0&e^{i\eta_2}&0 \\
0&0&1 \\
\end{pmatrix},
\end{align}
where $c_{ij}=\cos\theta_{ij}$, $s_{ij}=\sin\theta_{ij}$,
and $\delta_{\rm CP}$, $\eta_1$, and $\eta_2$ are imaginary phases, the regions for the allowed effective Majorana massses $\vert m_{ee}\vert$ and $\vert m_{\mu\mu}\vert$
are obtained with~\cite{Frigerio:2002rd}
\begin{align}
U_{e1}&= c_{12}c_{13}e^{i\eta_1},\\
U_{e2}&= s_{12}c_{13}e^{i\eta_2},\\
U_{e3}&= s_{13}e^{-i\delta_{\rm CP}},\\
U_{\mu1}&=-s_{12}c_{23}e^{i\eta_1}-c_{12}s_{13}s_{23}e^{i(\delta_{\rm CP}+\eta_1)},\\
U_{\mu2}&=c_{12}c_{23}e^{i\eta_2}-s_{12}s_{13}s_{23}e^{i(\delta_{\rm CP}+\eta_2)},\\
U_{\mu3}&=c_{13}s_{23}.
\end{align}

\begin{table}[!t]
  \centering
    \caption{For benchmark signal inputs of $\Lambda=200\TeV$ and $C_5^{\ell\ell'} = \delta_{\ell\mu}\delta_{\ell'\mu}$, the expected number of background and $\nuBB$ signal events in the signal region with $\mathcal{L}=300~\invfb$ (3 ab$^{-1}$).}
\resizebox{\columnwidth}{!}{
    \begin{tabular}{l|ccc|c|c}
    \hline    \hline
    Collider & QCD $W^\pm W^\pm j j$  & EW $W^\pm W^\pm j j$ &  $W^\pm V$ & Total & Signal\\ \hline
    LHC		& $<0.01$ &  6.40 &  1.16 &  7.56 & 0.013 \\
    HL-LHC	& $<0.01$ & 64.0 & 11.6 & 75.5 & 0.13\\
   \hline \hline
    \end{tabular}}
     \label{tab:totalRates}    
\end{table}

\bibliography{wwScattWeinberg_refs}
\end{document}